
\magnification=1200
\settabs 10 \columns
\def\k{k\!\!\!/}
\def\p{p\!\!\!/}
\def\eps{\epsilon}

\def\epsa{\epsilon^{\lambda_1}_{\alpha}(k_1)}
\def\epsb{\epsilon^{\lambda_2}_{\beta}(k_2)}
\def\epsc{\epsilon^{\lambda_3}_{\mu}(p_1)}
\def\epsd{\epsilon^{\lambda_4}_{\nu}(p_2)}
\def\la{\lambda_1}
\def\lb{\lambda_2}
\def\lc{\lambda_3}
\def\ld{\lambda_4}

\def\lalblcld{(\lambda_1,\lambda_2,\lambda_3,\lambda_4)}
\def\abcd{\alpha\beta\mu\nu}\magnification=1200
\noindent October 1993
\null\vskip 3truecm
\centerline{\titlefnt Testing the Higgs system at a photon-photon collider}
\vskip 3.5truecm
\centerline{H. Veltman}
\centerline{Service de Physique Th\'eorique, CEA-Saclay}
\centerline{F-91191 Gif-sur-Yvette Cedex, FRANCE}
\vskip 4.5truecm
\centerline{\bf Abstract}
\noindent
The level of sensitivity of the processes
$\gamma\gamma \to ZZ$, $\gamma\gamma \to W^+W^-$
and $\gamma\gamma \to t\bar t$ to the Higgs sector of the
Standard Model Lagrangian in the
energy region between 200 GeV and 1 TeV is examined.
The elementary Higgs boson is taken to have a mass less than 1 TeV.
Sizeable effects are found in the $ZZ$ and $t\bar t$ channels
if the incoming photons have the same helicity.
Also the possibility that the elementary Higgs boson does not exist
is examined. Assuming new physics to show up in the TeV energy
region the cross sections are evaluated according to the heavy
Higgs model.
For center of mass energy values close to 1 TeV interesting effects
are found
in the $t\bar t$ channel if the photons have the same helicity.
The limit of large Higgs mass is not unique. The parametrization of this
arbitrariness may be interpreted as a representation of the new physics.
The effects for the processes $\gamma\gamma\to ZZ$
and $\gamma\gamma\to t\bar t$ are investigated. These effects may
be correlated to a
possible resonance in $WW$ scattering in the TeV region.
\vfill
\eject
\noindent{\bf 1.\ \ Introduction}
\bigskip
Lately photon-photon ($\gamma\gamma$) physics has received
considerable attention.
This is due to the work by Ginzburg et. al. [1] and
others [2], who have shown that it
is possible to convert a high energy $e^+e^-$ collider into
a high energy $\gamma\gamma$ collider.
This is done through backscattering of low energy photons
off the initial $e^+(e^-)$ beam, where the backscattered
photon receives up to 80\% of the electron beam energy.
It is expected that the $\gamma\gamma$ collider, just like the
$e^+e^-$ collider, will be a facility that is able to do
precision measurements. An interesting feature is that the energy
distribution of the resulting $\gamma$ beam is increasingly monochromatic
with increasing energy. Accuracy may thus be better at higher energies.
At this stage, however, it is of course not exactly known what
accuracy can be achieved.

\bigskip
In the study of the Higgs sector of the Standard Model
$\gamma\gamma$ physics may prove to be a helpful tool,
as the elementary Higgs boson may be produced indirectly
through a vector boson loop or a top quark loop, see fig.1.
The Higgs subsequently decays into a pair of charged $W$
or neutral $Z$ bosons, or a top quark pair (assuming the Higgs mass is
sufficiently large).
The reaction of fig.1 with
initial electrons instead of photons gives a cross section
proportional to the ratio of electron mass
and vector boson mass, $m_e/M$, and is thus negligible.
Another example where $\gamma\gamma$
physics may be complementary to  $e^+e^-$ physics
is the study of anomalous tree level couplings. For instance
the $\gamma\gamma W^+W^-$ coupling may be tested
through the process $\gamma\gamma\to W^+W^-$ [3].
In this paper
we do not consider anomalous $\gamma$, $W^{\pm}$ and $Z$ couplings.

\bigskip
The Higgs sector of the Standard Model Lagrangian is needed to ensure
the renormalizability of the theory. As is well known there are
difficulties with respect to the cosmological constant. The model
requires the existence of an elementary Higgs boson, of which
there is sofar no experimental evidence. The Higgs sector may be
considered suspect, and perhaps the Higgs boson
as predicted by the Standard Model Lagrangian does not exist as such.
If this is indeed the case then as yet unknown new physics must exist
that takes the place of the Higgs system. According to studies of
longitudinally polarized vector boson scattering
new physics must show up in the TeV region [4-6]
if the mass of the elementary Higgs boson is not below 1 TeV.
For example QCD-like models predict for longitudinal
vector boson scattering the occurence of a resonance in the
isospin $I=1$ channel at around 2 TeV [7,8].
Analysis [9,10] of longitudinal vector boson scattering shows that
the occurrence of such a resonance is model dependent. It may be
parametrized [9] by means of the Lehmann $\beta$ parameter [11].
If $\beta$ could be measured in $\gamma\gamma$ processes then that could
be used to guess probable behaviour of the $WW$ scattering cross
section in the $I=1$ channel. Behaviour in other channels is dependent
on other parameters.

\bigskip
It is perhaps useful the explain the question of parametrization
in some detail. In the method of large Higgs mass the first parameter
occurring in the results is $\ln m^2$, where $m$ is the Higgs mass.
This logarithm occurs not only in $WW$ scattering but also in the
$\rho$-parameter and in $WW$ production by photons. It is of prime
interest if this parameter takes the same value in all those cases.
In the heavy Higgs model differences would have to come from yet
higher order terms, which in one study [12] have been shown to be
small if the Higgs mass is below 3 TeV. Use of an additional particle,
the $U$-particle [13], introduces a new parameter that may be
identified, within this model, with Lehmann's $\beta$ parameter.
Then possible resonance effects in $WW$ scattering may be related to
entirely different effects in $\gamma\gamma$ processes. Such effects
appear in a significant way for $t\bar t$ and longitudinal $ZZ$
final state processes.

\bigskip
The experiments that explore the energy region between 100 GeV and
1 TeV must identify one of the following possibilities:
\item{-} there is an elementary Higgs with a mass below 1 TeV;
\item{-} there is new physics.

\noindent The first possibility might by identified by direct Higgs search.
In the second case some model is required. Here we use as a model the
Standard Model Lagrangian in the limit of a heavy Higgs with an
additional Higgs interaction.

\bigskip
This paper is organized as follows. In section 2 we define the kinematics
and write the equation for the cross-section of the process
$\gamma\gamma\to X\bar X$, where $X\bar X$
represents a $ZZ$, a $W^+W^-$ or a $t\bar t$ pair.
The numerical values used for the various parameters are listed.
In section 3 the Higgs system is discussed in some detail.
In particular the heavy Higgs model and the introduction of an
additional interaction is described.
In section 4 we present a calculation of the cross section
for the process $\gamma\gamma\to ZZ$ to lowest non-zero order in
perturbation theory. We examine the cross section as a function of
the Higgs mass and compare the effects for the case of a low mass Higgs
with those for the case of our heavy Higgs model.
In section 5 we discuss the process $\gamma\gamma\to
W^+W^-$ and in section 6 the process $\gamma\gamma\to t\bar t$.
Section 7 contains a summary and a discussion of the results.
\bigskip
Our metric is such that $p^2=-m^2$ for an on mass-shell
 particle with mass $m$ and momentum $p$.
\bigskip
\bigskip
\noindent{\bf 2.\ \ Kinematics and definitions}
\bigskip
The process $\gamma(k_1)\gamma(k_2)\to X(p_1)\bar X(p_2)$
is displayed in fig.2. Here $X\bar X$ represents
the final $ZZ$, $W^+W^-$ or $t\bar t$ pair.
The momenta of the incoming photons $k_1$ and $k_2$
are defined to be aligned along the z-axis, thus
$$k_1=E(0,0,1,i),\ \ \ \ k_2=E(0,0,-1,i),\eqno(2.1)$$
with $E$ the photon beam energy.
The corresponding transverse circular polarization vectors are
defined by
$$\epsilon^+(k_1)=\epsilon^-(k_2)={1\over \sqrt{2}} (1,i,0,0),\ \ \ \
  \epsilon^-(k_1)=\epsilon^+(k_2)={1\over \sqrt{2}}
(1,-i,0,0).\eqno(2.2)$$
The polarization vectors are orthogonal to the momentum vectors,
thus
$$k_1\eps(k_1)=k_2 \eps(k_2)=0.\eqno(2.3)$$
In addition we have
$$k_1\eps(k_2)=k_2 \eps(k_1)=0.\eqno(2.4)$$
The $J=0$ state is for incoming
$\eps(k_1)^{\pm}\eps(k_2)^{\pm}$ polarization
states and the $J=2$ state is for incoming
$\eps(k_1)^{\pm}\eps(k_2)^{\mp}$ polarization states.
The momentum vectors $p_1$ and $p_2$
of the outgoing vector boson or top quark pair
are a function of the scattering angle $\theta$. We may write
$$p_1=E (\beta_x\sin\theta,0,\beta_x\cos\theta,i),
\ \ \ \ p_2=E (-\beta_x\sin\theta,0,-\beta_x\cos\theta,i),\eqno(2.5)$$
where the subscript $x$ denotes the particle considered.
For the neutral vector boson, the charged vector boson and the
top quark we have, respectively,
$$
\beta_z =\sqrt{1-{4M^2_0\over s}},
\ \ \ \ \beta_w =\sqrt{1-{4M^2\over s}},
\ \ \ \ \beta_t =\sqrt{1-{4m^2_t\over s}},
\eqno(2.6)$$
where as usual $s=4E^2$ is the center of mass energy squared.
If the final state is a pair of vector bosons we need to define
in addition to the two transverse polarization vectors $\eps^+$ and
$\eps^-$, the longitudinal polarization vector $\eps^0$:
$$\eqalignno{& \eps^+(p_1)=\eps^-(p_2)=
{1\over \sqrt{2}} (\cos\theta,-i,-\sin\theta,0),\cr
& \eps^-(p_1)=\eps^+(p_2)=
{1\over \sqrt{2}} (\cos\theta,i,-\sin\theta,0),\cr
&\eps^0(p_1)={E\over M_v}(\sin\theta,0,\cos\theta,i\beta_v),\cr
&\eps^0(p_2)={E\over M_v}(-\sin\theta,0,-\cos\theta,i\beta_v).
&(2.7)\cr}$$
The index $v$ represents the charged or neutral vector boson, with
$M_z=M_0$ and $M_w=M$.
The polarization vectors are orthogonal to the corresponding momentum
vectors,
thus $p_i\eps(p_i)=0$ for each $i=1,2$ with $\eps=\eps^{\pm},\eps^0$.
For a specific helicity configuration of the photons and vector bosons
the amplitude for $\gamma\gamma\to VV$, with $VV$ representing the $ZZ$ or
the $W^+W^-$ pair, may be written as
$$A^{vv}(\lambda_1,\lambda_2,\lambda_3,\lambda_4)=
\eps^{\lambda_1}_{\alpha}(k_1)\eps^{\lambda_2}_{\beta}(k_2)
\eps^{\lambda_3}_{\mu}(p_1)
\eps^{\lambda_4}_{\nu}(p_2)\cdot A^{vv}_{\alpha\beta\mu\nu}.\eqno(2.8)$$
There are altogether 36 different configurations, but not all of them
are independent. Due to the relations
$$A^{vv}(\lambda_1,\lambda_2,\lambda_3,\lambda_4)=
A^{vv}(-\lambda_1,-\lambda_2,-\lambda_3,-\lambda_4), \eqno(2.9)$$
and
$$|A^{vv}(\lambda_1,\lambda_2,\lambda_3,\lambda_4)|=
|A^{vv}(\lambda_2,\lambda_1,\lambda_4,\lambda_3)|,\eqno(2.10)$$
12 independent amplitudes remain (in fact the number of independent
amplitudes may be reduced to 8 by considering the replacement $p\to -p$).
The differential cross-section for each helicity configuration is
$${d\sigma^{vv}(\lambda_1\lambda_2\lambda_3\lambda_4)\over d\cos\theta}=
{\beta_v\over 32 \pi s}\cdot
|A^{vv}(\lambda_1,\lambda_2,\lambda_3,\lambda_4)|^2\cdot (389.352)
\ {\rm pb}\ \ ({\rm energy \ in \ TeV}).\ \  \eqno(2.11)$$
When the total cross-section is calculated
a factor 1/2 must be included for the process $\gamma\gamma\to ZZ$
because of Bose statistics.
Next we list the differential cross sections for the initial $J=0,2$
and final $TT,\ TL,\ LL$ helicity
configuration states ($T=$ transverse, $L=$ longitudinal):
\item{(1)} $J=0$ (initial $++=--$ state only)
$$\eqalignno{ {d\sigma^{vv}_{TT}\over d\cos\theta}
& ={d\over d\cos\theta}\{ \sigma^{vv}(++++)+2\sigma^{vv}(+++-)
+\sigma^{vv}(++--)\}\cr
{d\sigma^{vv}_{TL}\over d\cos\theta} & ={d\over d\cos\theta}\{
 2\sigma^{vv}(+++0)+2\sigma^{vv}(++-0)\}\cr
{ d\sigma^{vv}_{LL}\over d\cos\theta} & =
{d\sigma^{vv}(++00)\over d\cos\theta} &(2.12)\cr}$$
\item{(2)} $J=2$ (initial $+-=-+$ state only)
$$\eqalignno{
{d\sigma^{vv}_{TT}\over d\cos\theta} & =
{d\over d\cos\theta}\{ \sigma^{vv}(+-+-)+2\sigma^{vv}(+-++)+
\sigma^{vv}(+--+)\}\cr
{d\sigma^{vv}_{TL}\over d\cos\theta} & ={d\over d\cos\theta}\{
 2\sigma^{vv}(+-+0)+2\sigma^{vv}(+-0+)\}\cr
{d\sigma^{vv}_{LL}\over d\cos\theta} & ={d\sigma^{vv}(+-00)\over d\cos\theta}
&(2.13)\cr}$$

\noindent Note that we have used eqs.(2.9) and (2.10).

For the process $\gamma\gamma\to t\bar t$, we consider final state
unpolarized top quarks only and the amplitude is of the form:
$$A^{t\bar t}(\lambda_1,\lambda_2)=\eps^{\lambda_1}_{\alpha}(k_1)
\eps^{\lambda_2}_{\beta}(k_2)\cdot A^{t\bar t}_{\alpha\beta},\eqno(2.14)$$
with $A^{t\bar t}(+,+)=A^{t\bar t}(-,-)$
and $A^{t\bar t}(+,-)=A^{t\bar t}(-,+)$. The corresponding differential
cross-section for each helicity configuration is given by
$${d\sigma^{t\bar t}(\lambda_1\lambda_2)\over d\cos\theta}=
{N_c\beta_t\over 32\pi s}\cdot |A^{t\bar t}(\lambda_1,\lambda_2)|^2\cdot
(389.352) \ {\rm pb} \ \ ({\rm energy\ in \ TeV}).\eqno(2.15)$$
$N_c=3$ is the colour factor.
\bigskip
In the numerical evaluations presented in this paper, we
used the following values for the relevant parameters:
$$\eqalignno{
\alpha & ={e^2\over 4\pi}={1\over 128};\ \ \ \alpha_w={g^2\over 4\pi}
={1\over 30};\cr
M & =80.22\ {\rm GeV},\ {\rm charged\ vector\ boson\ mass\ };\cr
s_w^2 & =0.2326,\ {\rm sine\ squared\ of\ the\ weak\ mixing\ angle};\cr
M_0= & =91.173\ {\rm GeV},\ {\rm neutral\ vector\
boson\ mass\ }. & (2.16)\cr}$$
\bigskip
\bigskip
\noindent{\bf 3.\ \ Vector bosons and the Higgs}
\bigskip
\noindent{\bf 3.a\ \ The  conspiracy}
\bigskip
So far, up to an energy of 100 GeV, still nothing is known about the
Higgs sector in spite of the fact that the LEP experiments perform
their measurements with very high precision. This is a consequence
of the screening theorem [6]: when for a process the one loop correction
is calculated due to a heavy Higgs (but still with a mass less
than, say, 1 TeV), the dependence on the Higgs mass is only logarithmic.
Quadratic dependence shows up at the two-loop level. For example
the $\rho$ parameter including the heavy Higgs mass correction is
given by [12]
$$\eqalignno{\rho & = 1-{3\alpha\over 16\pi c_w^2}\cdot
{\rm ln}\left( {m^2\over M^2}\right) + 9.49\cdot10^{-4}
 \cdot {\alpha^2 \over s_w^2c_w^2} {m^2 \over M^2}\cr
& = 1-0.057\% \cdot
{\rm ln}\left( {m^2\over M^2}\right)+ 2.85\cdot 10^{-7}\cdot {m^2 \over M^2},
&(3.1) \cr}$$
where $m$ is the Higgs mass.
The experimental value is
$$\rho = 1\pm0.5\% .\eqno(3.2)$$
The error can still be reduced once more is known about the mass
of the top quark.
\bigskip
Experiments performed at a center of mass energy greater
that 170 GeV will be able to observe vector boson pair production.
This opens up new possibilities in the study of the Higgs system, since
the Higgs particle plays an important role concerning longitudinal
vector boson interaction. This is quite easy to see. In the high energy
limit the longitudinal polarization vector of eq.(2.7) is
proportional to $E/M_v$,
$$\eps^0(p)={p\over M_v}+{\cal O}\left( {M_v\over E} \right),\
M_v=M\ {\rm or}\ M_0,\eqno(3.3)$$
and for example the amplitude $A^{vv}(\lambda_1,\lambda_2,0,0)$ of eq.(2.8)
will seem to be proportional to $E^2/M_v^2$:
$$A^{vv}(\lambda_1,\lambda_2,0,0)=\eps^{\lambda_1}_{\alpha}(k_1)
\eps^{\lambda_2}_{\beta}(k_2)\cdot \left\{ {p_{1\mu}p_{2\nu}\over M_v^2}
\right\} \cdot A^{vv}_{\alpha\beta\mu\nu}.\eqno(3.4)$$
If physics is described according to the renormalizable Standard
Model Lagrangian containing a Higgs with a mass less
than 1 TeV then the amplitude for any process, like the
one above, behaves at
most as a constant in the large energy limit. Thus leading energy
cancellations must take place, which is indeed precisely what happens,
first of all due to the structure of the Yang-Mills vertices, but
also due to the Higgs interaction. Therefore if the Higgs does not
exist and if we consider the Standard Model Lagrangian in the limit
of a large Higgs mass, some leading energy terms may survive.
Thus studying the Higgs system may be done by studying longitudinally
polarized vector bosons. When we evaluate the heavy Higgs mass correction
for a typical process such as $e^+e^-\to W^+_LW^-_L$, the screening
theorem is still valid, but the $\ln m^2$ term is enhanced by a
factor $E^2/M^2_v$ [13-16]. We note here that the $\ln m^2$ term
should in fact
be interpreted as an unknown parameter (this issue will be
elaborated in section 3.c). It is expected that similarly enhanced
corrections occur
for $\gamma\gamma\to V_LV_L$.
\bigskip
At high energies such corrections may thus become very large, which
may seem very promising, were it not for the existence of the
transverse polarized vector boson $V_T$.
The transverse vector boson production may prove to be a severe background
to the longitudinal vector boson production for the following reasons:
\item{(1)} with respect to the Higgs system the $V_T$ behaves much like
the photon, i.e. it is rather insensitive to the Higgs system;
\item{(2)} although the decay products of the $V_L$ have a different
angular distribution than those of the $V_T$, still the $V_L$ can never
be fully isolated;
\item{(3)} $V_T$ production is always dominant over $V_L$ production,
the difference often being  a factor 10 or  more.

\noindent
There seems to exist a conspiracy to obstruct testing the Higgs system.
For energy values up to the vector boson mass  the only
heavy Higgs effects are proportional to $\ln m^2$.
While the screening theorem is still
valid, for energies well above the vector
boson mass the limit of a heavy Higgs mass leads to survival
of $E^2/M_v^2$ terms.
Other circumstances effectively  make detection of even these effects
very difficult.

Sensitivity to the Higgs system may thus not improve when the center
of mass energy is increased from the threshold value to 1 TeV.
As will be derived in section 5, the leading $\ln m^2$ correction
to the $\gamma \gamma\to W^+W^-$ cross section near threshold is
$$ < 1\% \cdot\ln m^2,\eqno(3.5)$$
depending on the renormalization scheme considered.

We now turn to the description of the Higgs
system for the special cases that are considered in this paper.
\bigskip
\noindent{\bf 3.b.\ \ Standard Model Higgs boson}
\bigskip
The existence of the Standard Model
elementary Higgs boson with a mass below 1 TeV is assumed.
An experimental lower bound of 60 GeV is provided by the LEP experiments.
Here however we assume that the Higgs is at least as heavy as twice
the vector boson mass.
We assume therefore the following mass range
$$2M < m < 1\ {\rm TeV}.\eqno(3.6)$$
Our aim is to examine the sensitivity of
 the processes $\gamma\gamma\to ZZ$ and $\gamma\gamma\to W^+W^-$
(the  $t\bar t$ channel will be discussed in section 6)
 to the Higgs exchange diagram of fig.1. Its contribution
may possibly be significant, notably in the resonance region,
where $-s+m^2={\cal O}(g^2)$. As we will see in section 4, for $\gamma
\gamma \to ZZ$ this is indeed the case for a relatively light Higgs.
For this process the diagram of fig.1 gives the following contribution
to the amplitude of eq.(2.8):
$$\eqalignno{A^{zz}_H(\lambda_1,\lambda_2,\lambda_3,\lambda_4)
& =\eps^{\lambda_1}_{\alpha}(k_1)
\eps^{\lambda_2}_{\beta}(k_2)
\eps^{\lambda_3}_{\mu}(p_1)
\eps^{\lambda_4}_{\nu}(p_2)\cr
& \cdot \delta_{\alpha\beta}\delta_{\mu\nu}  \cdot
P_H\cdot A(H)\cdot {\alpha\alpha_w\over c_w^2}.&(3.7)\cr}$$
For $\gamma\gamma\to W^+W^-$ we have
$$\eqalignno{A_H^{ww}(\lambda_1,\lambda_2,\lambda_3,\lambda_4)
& =\eps^{\lambda_1}_{\alpha}(k_1)
\eps^{\lambda_2}_{\beta}(k_2)
\eps^{\lambda_3}_{\mu}(p_1)
\eps^{\lambda_4}_{\nu}(p_2)\cr
& \cdot \delta_{\alpha\beta}\delta_{\mu\nu}
\cdot P_H\cdot A(H)\cdot\alpha\alpha_w,&(3.8)\cr}$$
where
$$\eqalignno{A(H) & = 6M^2+m^2 -{C_0^w\over i\pi^2}\cdot
(M^2m^2-7M^2s+12M^4)\cr
& + Q^2N_cm_t^2\cdot
\left(-4+{C_0^t\over i\pi^2}\cdot
(8m_t^2-2s)\right),&(3.9)\cr}$$
with as usual $N_c=3$ and $Q=2/3$ is the fractional charge of the top quark.
Note that, due to eqs.(2.3) and (2.4), the $k_{1\alpha}k_{2\beta}$
and $k_{1\beta}k_{2\alpha}$ pieces do not contribute and we therefore
did not bother to write them down. Only the $\delta_{\alpha\beta}$ piece
survives, which implies that the Higgs exchange graph of fig.1 will only
contribute when the incoming photons have the same helicity.
For the process $\gamma\gamma\to
W^+W^-$, the effect of the diagram of fig.1 has been investigated in ref.[17]
for Higgs mass values up to 300 GeV. The effect is small.
For higher Higgs mass values  the effect is completely negligible.
The reason is that, since it is a next to leading order effect,
it would only be significant if
 the size of the $W^+_LW^-_L$
cross section is of comparable order of magnitude to the size of
the $W^+_TW^-_T$
cross section. Such is not the case and except near threshold,
the longitudinal final state is highly
suppressed relative to the transverse final state.

The scalar three-point function $C^x_0$ appearing in eq.(3.9) is defined
by
$${C_0^x\over i\pi^2}=\int d_n q
{1\over {\{q^2+m_x^2\}}{\{(q+k_1)^2+m_x^2\}}{\{(q+k_1+k_2)^2+m_x^2\}}}.
\eqno(3.10)$$
The Higgs propagator $P_H$ is given by $1/(-s+m^2-i\eps)$. Near
the resonance we need to employ the Dyson summed propagator, i.e.
$$P_H={1\over -s+m^2-\Pi(s)},\eqno(3.11)$$
where $\Pi(s)$ is the collection of all irreducible self energy graphs
to all orders of perturbation theory. To first approximation
we may write
$$P_H={1\over -s+m^2-im\cdot \Gamma(g^2)},\eqno(3.12)$$
where $\Gamma(g^2)$ is the decay width of the Higgs boson evaluated
to lowest non-zero order in $g$.
Such an expansion, however, is only useful if perturbation theory
can be applied to the self energy $\Pi(s)$. For a heavy Higgs, matters become
rather ambiguous and for example for $m=1$ TeV the width is 0.5 TeV.
In ref.[18] this issue has been investigated for Higgs mass values
greater than 700 GeV and it has been suggested to employ
$\Gamma(s,g^2)$ in the propagator instead of $\Gamma(g^2)$, with
$$\Gamma(s,g^2)={s^2\over m^4}\cdot \Gamma(g^2).\eqno(3.13)$$
\bigskip
\noindent{\bf 3.c\ \  Heavy Higgs model}
\bigskip
If we consider the Standard Model in the limit of a large Higgs mass,
and if we
take this limit in the tree level Lagrangian we end up with an effective
field theory where the Higgs sector is given by the non-linear
$\sigma$-model. According to this effective theory, the tree amplitude
for $W_LW_L$ scattering grows like the energy squared and the unitarity
limit is reached at about 1 TeV. Thus for this  theory,  beyond 1 TeV,
physics becomes non-perturbative and we do not know how to calculate
things anymore. In the TeV region new physics will show up,
but what kind of new physics is at this moment anybody's guess.
At the same time we do know that up to 100 GeV the Standard
Model works extremely well, Higgs or no Higgs. In other words, given
the fact that there exists new physics, the Standard Model Lagrangian in the
heavy Higgs mass limit is a very good approximation.
Here we assume that up to 1 TeV the one-loop approximation is still
reasonable. If the non-linear $\sigma$-model is used for the Higgs sector,
then after performing a  typical one-loop calculation
infinities remain, i.e. terms containing
        $$-{2\over n-4}=\Delta,\eqno(3.14)$$
if we use dimensional regularization. They need to be interpreted
as arbitrary parameters which must be fixed by experiment.
Sensitivity to these unknown
parameters implies sensitivity to the, still unspecified, new physics.
\bigskip
If we consider the renormalizable Standard Model Lagrangian with the linear
$\sigma$-model as the Higgs sector and take the heavy Higgs mass limit at
the end of a one-loop calculation no infinities survive. Instead
we are left with terms proportional to $\ln m^2$. There is a direct
correspondence between these $\ln m^2$ terms and the $\Delta$ terms [19]:
$$\Delta \leftrightarrow \ln\left( {m^2\over M^2}\right),\eqno(3.15)$$
and thus the $\ln m^2$ terms are to be interpreted as unknown
parameters.
Since the one loop dependence on the Higgs mass is given by the $\ln m^2$
term only, one would expect the heavy Higgs model to contain just one
arbitrary parameter. However there is at least one more.
Take the amplitude for $W_LW_L$ scattering to one loop
order accuracy. Evaluated in the energy region $M\ll p\ll m$,
where $p$ is a typical momentum, it will be of the form
$$\eqalignno{A(W_LW_L\to W_LW_L) & =
\alpha_w\cdot{p^2\over M^2}\cdot c_1\cr
& +\alpha_w^2\cdot{p^4\over M^4}\cdot
\left\{ c_2\cdot \ln {m^2\over p^2} + \beta+{\cal O}\left({M^2\over p^2}
\right) \right\}+{\cal O}\left( {p^2\over m^2}\right).
\ \ \ \ \ \ &(3.16)\cr}$$
Here $c_1,\ c_2$ and $\beta$ are of the order one and are found by explicit
calculation. Besides the $\ln m^2$ term, the subleading term $\beta$
also needs to be interpreted as an unknown parameter [9]. The reason is that
the limit of a heavy Higgs mass is not unique. Taking the limit of a
heavy Higgs mass in the Lagrangian (non-linear $\sigma$-model) gives
$\beta=1/3$, while taking that limit after computing the one loop graphs
gives $\beta=-0.32$.

Evaluated in the same energy region, the amplitudes for the processes
$\gamma\gamma\to W^+_LW^-_L$ and $\gamma\gamma\to Z_LZ_L$ are of the form
$$\eqalignno{A^{ww}(+,+,0,0) & =\alpha\cdot{\cal O}\left( {M^2\over p^2}
\right)\cr
& +\alpha\cdot\left\{ \alpha_w\cdot {p^2\over M^2}\cdot c_1^w+
\alpha_w^2\cdot{p^4\over M^4}\cdot
 \left\{ c_2^w\cdot \ln {m^2\over p^2} +c^w_3+\ldots
\right\}\right\},\cr
A^{zz}(+,+,0,0) & =
\alpha\cdot\left\{ \alpha_w\cdot {p^2\over M^2}\cdot c_1^z+
\alpha_w^2\cdot{p^4\over M^4}\cdot
 \left\{ c_2^z\cdot \ln {m^2\over p^2}+c^z_3+\ldots
\right\}\right\}.\ \ \ \ \ \ &(3.17)\cr}$$
The photons need to have the same helicity; the amplitudes for
which the photons have the opposite helicity do not contain
terms that lead to bad high energy behaviour in the heavy Higgs mass limit.

The terms $c_2^w$ and $c_2^z$ are found by explicit calculation.
In the heavy Higgs model the parameters $\ln m^2$, $c_3^w$ and $c_3^z$
are interpreted as unknown parameters. Comparing the expansions of
eqs.(3.16) and (3.17) we can relate the
$\ln m^2$ terms, while the connection between $\beta$, $c_3^w$ and
$c_3^z$ seems to be completely lost. This would sugggest
that for every process there is  an extra arbitrary parameter.
However, as we will demonstrate in the next section,
these parameters may be uniquely related to each other.
Thus assuming for $W_LW_L$ scattering the one loop approximation
to be a reasonable one,
the heavy Higgs model contains just two arbitrary parameters.
We note that if this assumption is incorrect and if
higher order corrections are important,
the expansions of eqs.(3.16) and (3.17)
cannot be related to each other; the set of parameters needed to
describe one process may be  different from the set of
parameters for another process.
\bigskip
Consider first the process $\gamma\gamma\to ZZ$ described
according to the heavy Higgs mass model.
The Higgs exchange diagram of fig.1 contributes to the lowest non-zero
order in perturbation theory and from eqs.(3.7), (3.9) and (3.11)
we find in the limit of large $m$:
$$\eqalignno{
\lim_{m \to \infty} A_H^{zz}(\lambda_1,\lambda_2,\lambda_3,\lambda_4)
 & = \eps^{\lambda_1}_{\alpha}(k_1)\eps^{\lambda_2}_{\beta}(k_2)
\eps^{\lambda_3}_{\mu}(p_1) \eps^{\lambda_4}_{\nu}(p_2)\cr
\cdot \delta_{\alpha\beta}\delta_{\mu\nu}
& \cdot  {\alpha\alpha_w\over  c_w^2}\cdot
\left\{ 1-{C_0^w\over i\pi^2}\cdot
M^2+{\cal O}\left( {s\over m^2},{M^2\over m^2}
\right)\right\}. &(3.18)\cr}$$
Just like in $W_LW_L$ scattering, the amplitude for the process
$\gamma\gamma\to Z_LZ_L$ shows bad high energy behaviour already in lowest
non-zero order. In the expansion of eq.(3.17) $p^2=-s$ with $c_1^z=1/2$.
The $\ln m^2$ terms will arise at the next to
leading order, in this case thus at the two-loop level.

Unfortunately, unlike in $WW$ scattering, $\gamma\gamma\to ZZ$
is the perfect example of the conspiracy described in section 3.a:
the $Z_TZ_T$ production is about a factor 100 larger, leaving the bad
high energy behaviour of the $Z_LZ_L$ amplitude almost undetectable.
The difference between a 1 TeV Higgs and an infinitely heavy Higgs
is initially an enhancement of about a factor 2-4
(depending on the top quark mass) for the $Z_LZ_L$
cross-section at $\sqrt{s}=1$ TeV. When transverse vector boson production is
included, the enhancement is reduced to a few percent. Maybe such effects
could be seen but now we are dealing with a theoretical uncertainty:
first of all due to the broad Higgs width, but also if we need an
accuracy of a few procent then next to leading order calculations
must be done. This would entail a two-loop calculation.
\bigskip
For the process $\gamma\gamma\to W^+W^-$ the background due to the
transverse vector boson production is even more substantial. To
order $\alpha$, the $W^+_TW^-_T$ amplitude is a constant in the
large energy limit while the $W_L^+W_L^-$ amplitude vanishes like
$M^2/s$. The quadratic energy term, that
arises at order $\alpha\alpha_w$,
is thus highly suppressed and heavy Higgs effects are not expected to
be substantially different at threshold  then at $\sqrt{s}=1$ TeV.
Effects only become interesting for center of mass energy values above
1.5 TeV.

It may be interesting to know what the leading Higgs
mass correction will be near threshold.
Making an expansion in the Higgs mass,  with $M\simeq p\ll m$, the amplitude
to one loop order accuracy is of the form
$$A(\gamma\gamma\to W^+W^-)=\alpha\cdot c_a^w+\alpha\alpha_w
\cdot\left\{
 c_b^w\cdot\ln {m^2\over M^2}+{\cal O}\left( 1 \right)\right\},
\eqno(3.19)$$
for any helicity configuration of the photons and vector bosons.
The parameters $c_a^w$ and $c^w_b$ are found by explicit calculation.
The one loop calculations that must be done in order to extract
the $\ln m^2$ term, i.e. the parameter $c^w_b$, are given in full detail
elsewhere [12,15].
In the end we only need to consider the one-loop corrected vertices,
while the propagators remain unchanged
(except for the Higgs propagator). After that only irreducible box diagrams
must be calculated. In ref.[15] the leading $\ln m^2$ corrections
are listed for a number of vertices.

Concerning finite renormalization, counterterms
are fixed such that the masses are located at the pole of the propagators.
This fixes $\delta_M,\ \delta_{s_w},\ \delta_{m_t}$, and $\delta_m$
when making the shifts $M\to M(1+\delta_M)$, etc.
There is still one more parameter in the Standard Model tree level
Lagrangian,
namely the coupling constant $g$. Therefore one more measuring point is
needed, thereby obtaining the corresponding value for
$\delta_g$. For example when choosing  muon decay, we find:
$$\delta_g({\rm muon\ decay})={\alpha_w\over 4\pi}\cdot
{1\over 24}\cdot \ln m^2.
\eqno(3.20)$$
We also could have chosen the cross section for Coulomb scattering
at zero momentum transfer as our measuring point. We obtain
a different value for $\delta_g$:
$$\delta_g({\rm Coulomb\ scattering})=  {\alpha_w\over 4\pi}
\cdot {-5\over 12}\cdot\ln m^2.\eqno(3.21)$$
The correction due to $\delta_g$ is always proportional to the
amplitude in lowest non-zero order and will therefore not
induce bad high energy behaviour.
\bigskip
\noindent{\bf 3.d\ \ The $U$ particle}
\bigskip
The heavy Higgs mass limit is arbitrary.
Some time ago the $U$ particle has been introduced to parametrize this,
and to make this arbitrariness explicit [13].
To the $U$ particle corresponds a scalar singlet field, added to the
Lagrangian in a gauge invariant way. The $U$-particle, not coupled to the
vector bosons, is coupled to the Higgs with a strength $m^2$ and its mass
is taken to be equal to the Higgs mass:
$${\cal L}(U) = -{1\over 2}(\partial_{\alpha} U)^2
-{1\over 2}m^2 U^2-{gg_Um^2\over 4M}U^2H
-{g^2g_U^2m^2\over 4M^2}U^2(H^2+\phi^2)\, .\eqno(3.22)$$
The field $\phi$ is the usual Higgs ghost.
Since the $U$ is coupled only to the Higgs,
no dependence on $g_U$ should remain in the limit $m\to \infty$,
if the heavy Higgs mass limit is unique. This is not the case, and
the contribution of the $U$-particle to the amplitude of eq.(3.16)
is given by
$$\beta\to \beta+\beta^U,\eqno(3.23)$$
with
$$\beta^U=g^2_U\cdot\left( {\pi\over \sqrt{3}}-2 \right),\eqno(3.24)$$
thus demonstrating the arbitrariness of $\beta$.
It is very interesting to note that within the framework of
partial wave analysis the location of a resonance in the isospin $I=1$
channel for $W_LW_L$ scattering precisely depends on this $\beta$ term [9].
When $\beta={1 \over 3}$ the resonance is located at 9 TeV, while
if $\beta\leq 0$ no resonance will occur. For $\beta=5$ the
resonance occurs at 2 TeV, much like the scaled up version of the
$\rho$ resonance in $\pi\pi$ scattering.
Thus, within the framework of partial wave analysis, the experimental
sensitivity below 1 TeV to the $\beta$ parameter will enable us to
predict at what energy in the TeV region a resonance will occur in the
$I=1$ channel, however not explaining as to what kind of new physics
could cause such a resonance.
\bigskip
The $U$ particle effect for the $\rho$ parameter has been calculated
in ref.[20]; there the effect was found to be too small to be of
practical interest.
It may be quite interesting to see what an effect the existence
of such a resonance would have for the process $\gamma\gamma\to ZZ$.
The $U$ particle contribution to the
diagram of fig.1 is given by
$$\eqalignno{ & A_U^{zz}(\lambda_1,\lambda_2,\lambda_3,\lambda_4)=
\eps^{\lambda_1}_{\alpha}(k_1) \eps^{\lambda_2}_{\beta}(k_2)
\eps^{\lambda_3}_{\mu}(p_1) \eps^{\lambda_4}_{\nu}(p_2)\cr
& \cdot \delta_{\alpha\beta}\delta_{\mu\nu}\cdot
{\alpha\alpha_w^2\over  4\pi c_w^2} \cdot
\left( - {\beta^U\over 8M^2} \right)
\cdot \left\{ 6M^2+s -{C_0^w\over i\pi^2}6M^2(2M^2-s)
\right. \cr
& \left. +Q^2N_cm_t^2\cdot \left( -4m_t^2+{C_0^t\over i\pi^2}
(8m_t^2-2s)\right) \right\}.&(3.25)\cr}$$
When the initial photons have the same helicity and the outgoing
vector bosons are longitudinally polarized, the contribution to
the parameter $c^z_3$ of eq.(3.17) is given by
$$c^z_3\to c^z_3+c_3^{zU},\eqno(3.26)$$
with
$$c^{zU}_3={\beta^U\over 64\pi}.\eqno(3.27)$$
For $\beta=5$ this implies a $-50$\% effect in the cross section
for the process $\gamma\gamma\to Z_LZ_L$ at $\sqrt{s}=1$ TeV.
But as we remarked before, such an effect is unfortunately probably
undetectable when transverse polarized vector bosons are included
in the final state: it is then reduced by a factor $10^{-2}$ to $-0.5$\%.
\bigskip
\bigskip
\noindent{\bf 4.\ \ The process $\gamma\gamma\to ZZ$}
\bigskip
The evaluation of the cross-section in lowest non-zero order in
perturbation theory has been done by Jikia [21], who found that
the transverse vector boson production is overwelmingly large as
compared to the longitudinal vector boson production. Since at first
it appears that $\gamma\gamma\to Z_LZ_L$ is an ideal place to test
for the Higgs sector, this result is disappointing; perhaps not
so surprising in view of the conspiracy described in section 3.a.

Nevertheless such a result needs to be verified, not in the least since
the corresponding calculation is quite complicated. While preparing
this work we became aware of two more independent calculations, performed
by Berger [22] and Dicus and Kao [23].
Our results confirm all of these three independent calculations.

Box diagrams need to be calculated, for which there is also a fermionic
contribution. This contribution has been calculated in ref.[24] for
the gluon fusion process, i.e. for the process $gg\to ZZ$. This has
provided us with an additional check, and we found agreement here as well.
\bigskip
We performed our calculations in the Feynman-'t Hooft gauge.
\bigskip
\noindent{\bf 4.a\ \ The amplitude}
\bigskip
The diagrams that contribute to the amplitude for $\gamma\gamma\to ZZ$
to lowest non-zero order in perturbation theory are of order $g^4$,
and are shown in fig.3. For the amplitude of eq.(2.8),
$$A^{zz}(\lambda_1,\lambda_2,\lambda_3,\lambda_4)  =
\eps^{\lambda_1}_{\alpha}(k_1)\eps^{\lambda_2}_{\beta}(k_2)
\eps^{\lambda_3}_{\mu}(p_1)\eps^{\lambda_4}_{\nu}(p_2)\cdot
A^{zz}_{\abcd},\eqno(4.1)$$
the expression for $A^{zz}_{\abcd}$ takes the form
$$\eqalignno{A^{zz}_{\abcd}=
& \{ k_{1\mu}k_{1\nu}p_{1\alpha}p_{1\beta}
+    k_{2\mu}k_{2\nu}p_{1\alpha}p_{1\beta}\}\cdot f_1\cr
+ &  k_{1\mu}k_{2\nu}p_{1\alpha}p_{1\beta}\cdot f_2
+    k_{2\mu}k_{1\nu}p_{1\alpha}p_{1\beta}\cdot f_3\cr
+ & \delta_{\mu\nu}p_{1\alpha}p_{1\beta}\cdot f_4
+ \{ \delta_{\mu\alpha}k_{1\nu}p_{1\beta}-
  \delta_{\nu\beta}p_{1\alpha}k_{2\mu}\}\cdot f_5\cr
+ & \{ \delta_{\mu\alpha}k_{2\nu}p_{1\beta}-
  \delta_{\nu\beta}p_{1\alpha}k_{1\mu}\}\cdot f_6
+ \{ \delta_{\mu\beta}p_{1\alpha}k_{1\nu}-
  \delta_{\nu\alpha}k_{2\mu}p_{1\beta} \}\cdot f_7\cr
+ & \{ \delta_{\mu\beta}p_{1\alpha}k_{2\nu}-
  \delta_{\nu\alpha}k_{1\mu}p_{1\beta}\}\cdot f_8
+ \{ \delta_{\alpha\beta}k_{1\mu}k_{1\nu}+
  \delta_{\alpha\beta}k_{2\mu}k_{2\nu}\}\cdot f_9\cr
+ & \delta_{\alpha\beta}k_{1\mu}k_{2\nu}\cdot f_{10}
+ \delta_{\alpha\beta}k_{1\nu}k_{2\mu}\cdot f_{11}\cr
+ & \delta_{\alpha\beta}\delta_{\mu\nu}\cdot f_{12}
+   \delta_{\alpha\mu}\delta_{\beta\nu}\cdot f_{13}
+   \delta_{\alpha\nu}\delta_{\beta\mu}\cdot f_{14}. &(4.2)\cr}$$
We used momentum conservation, and substituted $p_2=k_1+k_2-p_1$.
As $A^{zz}_{\abcd}$ is to be multiplied with the
polarization vectors, the terms proportional to $k_{1\alpha}$,
$k_{1\beta}$, $k_{2\alpha}$, $k_{2\beta}$, $p_{1\mu}$ and $p_{2\nu}$
do not contribute here and we left them out.
In a Ward identity one will encounter such amplitudes multiplied
by some momentum instead of one or more of the polarization vectors.
Then terms containing $k_{1\beta}$ etc. must be kept.

The algebraic expressions for the functions $f_1,\ldots,f_{14}$ were
derived using Schoonschip [25], and subsequently numerically
evaluated using Formf [26].

Referring to section 3, it may be deduced easily that
the diagram of fig.1 contributes to the $f_{12}$ term only.
For the case that the Higgs exists with a mass
less than 1 TeV it is given by
$$\eqalignno{f_{12H} & ={\alpha\alpha_w\over c_w^2}
\cdot {1\over -s+m^2-im\cdot \Gamma(s,g^2)}\cr
& \cdot\left\{ 6M^2+m^2-{C^w_0\over i\pi^2}\cdot
(M^2m^2-7M^2s+12M^4)\right. \cr
& + \left. Q^2N_cm_t\cdot\left(
-4+{C_0^t\over i\pi^2}\cdot (8m^2_t-2s)\right)
\right\}&(4.3)\cr}$$
This expression differs slightly from the one given in [21], but the
numerical consequences are negligible.
For a very heavy Higgs we have,
$$\lim_{m\to \infty}f_{12H}={\alpha\alpha_w\over c^2_w}
\cdot \left\{ 1-{C^w_0\over i \pi^2}\cdot M^2\right\}\to
{\alpha\alpha_w\over c^2_w}\cdot \{ 1\}{\rm \ \ for\ large\ }s.\eqno(4.4)$$
The $U$-particle contribution is given by:
$$\eqalignno{f_{12U} & ={\alpha\alpha_w^2\over 4 \pi c_w^2}
\cdot \left( -{\beta^U\over 8M^2}\right)
\cdot\left\{ 6M^2+s-{C^w_0\over i\pi^2}\cdot 6M^2\cdot (2M^2-s)\right. \cr
& \left. +Q^2N_cm_t^2\cdot\left(
-4+{C_0^t\over i\pi^2}\cdot (8m^2_t-2s)\right)
\right\}\to {\alpha\alpha_w\over c_w^2}\cdot
\left\{ -{\alpha_w \beta^U s\over 32\pi M^2}\right\}{\rm \ \ for \ large \ s}
.\ \ \ &(4.5)\cr}$$
To order $g^4$ all other functions $f_i$ are insensitive to the
Higgs system.
If $\beta^U=5$ the $U$ particle contribution leads to a $-25\%$ effect
to the term $f_{12H}$ of eq.(4.4) at $\sqrt{s}=1$ TeV, which corresponds
to a $-50\%$ effect for the $Z_LZ_L$ cross section.
As we mentioned before, in the process
$\gamma\gamma\to ZZ$ such an effect is unfortunately reduced by two orders
of magnitude due to the enormous $Z_TZ_T$ background.
\bigskip
\noindent{\bf 4.b\ \ Ward identities; the equivalence theorem}
\bigskip
In order to verify the correctness of the functions $f_1,\dots,f_{14}$
we must consider the relevant Ward identities of which there are
six altogether, which may be written as
$$(ik_{1\alpha})\cdot\epsb\epsc\epsd\cdot A^{zz}_{\abcd}=0,\eqno(4.6)$$
$$(ik_{2\beta})\cdot\epsa\epsc\epsd\cdot A^{zz}_{\abcd}=0,\eqno(4.7)$$
$$(ik_{1\alpha}ik_{2\beta})\cdot\epsc\epsd\cdot A^{zz}_{\abcd}=0,\eqno(4.8)$$
$$\eqalignno{& ( -ip_{1\mu})\cdot\epsa\epsb\epsd\cdot A^{zz}_{\abcd}\cr
+ & M_0\cdot \epsa\epsb\epsd\cdot A^{\phi z}_{\alpha\beta\nu}=0,&(4.9)\cr}$$
$$\eqalignno{& ( -ip_{2\nu})\cdot\epsa\epsb\epsc\cdot A^{zz}_{\abcd}\cr
+& M_0\cdot \epsa\epsb\epsc\cdot A^{z\phi}_{\alpha\beta\mu}=0,&(4.10)\cr}$$
$$\eqalignno{ & (ip_{1\mu}ip_{2\nu})\cdot\epsa\epsb\cdot A^{zz}_{\abcd}\cr
+ & M_0\cdot (-ip_{2\nu})\epsa\epsb\cdot A^{\phi z}_{\alpha\beta\nu}\cr
+ & M_0\cdot (-ip_{1\mu})\epsa\epsb\cdot A^{z\phi}_{\alpha\beta\mu}\cr
+ & M_0^2\cdot \epsa\epsb\cdot A^{\phi\phi}_{\alpha\beta}=0.& (4.11)\cr}$$
The first three Ward identities are also known as transversality conditions,
i.e. the amplitude equals zero when the photon polarization vector
is replaced by the corresponding momentum vector. The other three Ward
identities involve the amplitude for which the polarization vector of
the massive vector boson is replaced  by its momentum vector.
$A^{\phi Z}(\la,\lb,\ld)
=\epsa\epsb\epsd A^{\phi z}_{\alpha\beta\nu}$ denotes the
amplitude for $\gamma\gamma\to \phi Z$, obtained from $\gamma\gamma\to
ZZ$ with the replacement of $Z_{\mu}$ by the Higgs ghost $\phi$.
Similarly, $A^{Z\phi} (\la,\lb,\lc)$ and $A^{\phi\phi}(\la,\lb)$ correspond
to, respectively, the $\gamma\gamma\to Z\phi$ amplitude and the
$\gamma\gamma\to \phi\phi$ amplitude. In Ward identities momenta are
defined to be flowing into the vertex, which leads to the extra minus
signs appearing in eqs.(4.9), (4.10) and (4.11).

As an illustration, consider the Ward identity of eq.(4.6) for the
$k_{1\mu}k_{1\nu}p_{1\beta}$ term. After multiplication of $A^{zz}_{\abcd}$
of eq.(4.2) with $k_{1\alpha}$, we obtain for this  term:
$$ \{ (p_1k_1)\cdot f_1+f_5-f_9\}
\cdot k_{1\mu}k_{1\nu}p_{1\beta}.\eqno(4.12)$$
There is one additional term, which we did not explicitly write down;
before multiplication with the
external polarization or momentum vectors it  is of the form
$k_{1\mu}k_{1\nu}k_{2\alpha}p_{1\beta}\cdot g_1$. This term does
not contribute to $A^{zz}(\la,\lb,\lc,\ld)$ because $\eps_1k_2=0$. However
it does contribute to the Ward identity of eq.(4.6) since
$k_1k_2=k^2/2=-s/2$ is not zero. The Ward identity is then given by
$$(p_1k_1)\cdot f_1+f_5-f_9-{1\over 2}s\cdot g_1=0.\eqno(4.13)$$
There are of course many more such identities.
\bigskip
\noindent From the three Ward identities of eqs.(4.9), (4.10), and (4.11)
we can derive the equivalence theorem [8,27], valid in the limit $M_0\ll E$.
The equivalence theorem states that the leading energy term of the
$Z_LZ_L$ amplitude is given by the amplitude where the
longitudinal vector boson  $Z_L$ is replaced by the corresponding
Higgs ghosts $\phi$. Here follows a sketch of the derivation.
In the large energy limit, the longitudinal polarization vector
$\eps^0_{\mu}(p)$ may be written as
$$\eps^0_{\mu}(p)={p_{\mu}\over M_0}+v_{\mu},\ \
v_{\mu}={\cal O}\left( {M_0\over E} \right).\eqno(4.14)$$
Using this relation,
the first step is to eliminate the momentum vectors appearing in the
Ward identities. This leads to identities
involving the longitudinal polarization vectors $\eps^0$, and the
vectors $v$. Next, substracting the identities (4.9) and (4.10)
from (4.11) leads to the relation
$$\eqalignno{ A^{zz}(\la,\lb,0,0) & = i^2\cdot A^{\phi\phi}(\la,\lb)\cr
& -i\cdot v_{1\mu}\epsa\epsb\cdot A^{\phi z}_{\alpha\beta\mu}\cr
& -i\cdot v_{2\nu}\epsa\epsb\cdot A^{z\phi}_{\alpha\beta\nu}\cr
&  + v_{1\mu}v_{2\nu}\epsa\epsb\cdot A^{zz}_{\alpha\beta\mu\nu}.&(4.15)\cr}$$
In the limit that $v$ is small, we obtain the equivalence
theorem:
$$A^{zz}(\la,\lb, 0,0)=-A^{\phi\phi}(\la,\lb)+{\cal O}\left( {M_0\over E}
\right)\eqno(4.16)$$
There do exist cases where it is not justified to throw away the diagrams
that are multiplied with one or more $v$'s.
Using the equivalence theorem in the large Higgs mass limit, we find
$$A^{zz}(+,+,0,0)=-{\alpha \alpha_w}\cdot {s\over 2 M^2},
\ \ M_0 \ll E \ll m,\eqno(4.17)$$
which may be compared to the term $f_{12H}$ of eq.(4.4).
The equivalence theorem is thus in fact a Ward identity for the
leading energy term only and it serves as another check on the
calculation.
We remark that in this case calculating the $\phi\phi$-amplitude is
by far simpler than calculating the $Z_L Z_L$-amplitude.

Using the theorem, the process $\gamma\gamma\to Z_LZ_L$ has been
studied in refs. [28,29] for center of mass energy values in the TeV range.
\bigskip
\noindent{\bf 4.c\ \ Results}
\bigskip
In fig.4a we show the $J=0$
differential cross-sections at $\sqrt{s}=200$ GeV.
When we take the limit
$m\to\infty$, the cross section is less than 10 fb. However if the Higgs
exists, and if $\sqrt{s}$ is the resonance energy value, i.e. $m=200$ GeV,
the cross section is enhanced by almost a factor $10^3$ and is 7 pb.
\bigskip
It is of course possible that the center of
mass energy is not the resonance energy value. In that case
a tail of a resonance could be observed for energy values
not too far away from the Higgs mass value. In order to have
an indication of the size of the effect of such a tail we introduce
the function $R(m,\theta,\sqrt{s})$. It is the ratio of
the differential cross-section for a given Higgs mass $m$ and the
differential cross-section in the limit of a large Higgs mass,  at the
center of mass energy $\sqrt{s}$ considered. In percentage
value:
$$R(m,\theta,\sqrt{s})=\left( {d\sigma(\theta)_m\over
d\sigma(\theta)_{m\to\infty}}-1\right) \cdot 100\%,\ \ \
d\sigma(\theta)={d\sigma\over d\cos\theta}.\eqno(4.18)$$
As usual the dependence of the differential cross section on $\sqrt{s}$
is understood and needs not be indicated.
$R(m,\theta,\sqrt{s})$ measures a lowest non-zero order
effect for the process $\gamma\gamma\to ZZ$. Hence the $U$ particle effect,
to be discussed at the end of this section, is not included.
$R(m,\theta,\sqrt{s})$ is positive for $m >\sqrt{s}$,
negative for $m < \sqrt{s}$.
In fig.4b we show $R(m,\theta,\sqrt{s}=200{\rm \ GeV})$
for 4 different Higgs mass values, namely $m=300$,
$400$, $500$, and $600$ GeV.

In fig.4c we show the $J=2$ differential cross sections which
to ${\cal O}(g^4)$ are insensitive to the Higgs  system.

In all cases,
varying the top quark mass from 120 to 180 GeV leads to
an insignificant effect.
\bigskip
Figs. 5a-5c show similar curves, but here at $\sqrt{s}=300$ GeV.
For the case
$m\to \infty$, dependence on the top quark mass $m_t$ is negligible.
The differential cross sections are of the
order of 0.1 pb and already much larger than at $\sqrt{s}=200$ GeV.
This is because the cross section for transverse vector boson production
increases rather steeply, and is in fact near its maximum value.
The longitudinal cross-section remains  in the fb range.

The dependence on $m_t$ is significant when $m=300$ GeV. The reason is
that for $\sqrt{s} > 2m_t$, the imaginary part of the top quark loop in the
$\gamma\gamma H$ vertex is non-zero and may lead to a large effect.
In fig. 5a the $J=0$ differential cross sections, with unpolarized
vector bosons in the final state, are shown
for top quark mass values of 120 GeV and 180 GeV.

In fig.5b $R(m,\theta,\sqrt{s}=300\ {\rm GeV})$ is plotted for
Higgs mass values $m=$ 400, 500 and 600 GeV and
for top quark mass values $m_t=120$ and 180 GeV.
\bigskip
Figs.6a-6c display the curves at $\sqrt{s}=500$ GeV. At this energy, given
a Higgs mass of 500 GeV, the resonance effect has disappeared.
Nevertheless, for small values of $|\cos\theta|$, there is a 10-15\%
effect with respect to the cross-section for
$m\to \infty$ for $m_t=120$ GeV. The effect is 5-8\% for $m_t=180$ GeV.
\bigskip
Finally, figs.7a-7d give the cross-sections at $\sqrt{s}=1$ TeV. Effects
of a Higgs with a mass of 1 TeV are at most 5\%
for $m_t=120$ GeV and at most 1.6\% for $m_t=180$ GeV.
\bigskip
We have also calculated  the $U$-particle effect in the limit of a
large Higgs mass, assuming a 2 TeV resonance in the
$I=1$ channel for $W_LW_L$
scattering. In fig.8 we show the $U$ particle effect to the $J=0$
differential cross-section in percentage value.
For the $LL$ final state the effect is always increasing
with increasing energy:
$-0.7$\% at $\sqrt{s}=200$ GeV and $-50$\% at $\sqrt{s}
=1$ TeV. However, when the $TT$ final state is included the effect is
always of the order
$-(0.1-0.4)\%$ and thus extremely difficult to observe. We note that effects
only become substantial for $\sqrt{s}>1.5$ TeV.
\bigskip
\bigskip
\noindent{\bf 5.\ \ The process $\gamma\gamma\to W^+W^-$}
\bigskip
Regarding the anomalous tree level $\gamma WW$ and $\gamma\gamma
WW$ couplings the process $\gamma\gamma\to WW$ has been discussed
extensively in the literature [3]. Here we do not consider such effects,
and we assume the tree level couplings to be the Yang-Mills couplings
as predicted by the Standard Model Lagrangian.

If the initial photons have the same helicity the $W^+_LW^-_L$
channel is highly supressed compared to the $W^+_TW^-_T$ channel.
Therefore if we consider unpolarized vector bosons in the final
state the Higgs exchange diagram of fig.1 plays little or no role, even for
center of mass energy values near the Higgs resonance [17].
In the case that the Higgs does not exist and if we consider the
heavy Higgs model, leading energy terms survive. For the process
$\gamma\gamma\to W^+_LW^-_L$ this happens only when the incoming
photons have the same helicity. This leading energy term, corresponding
to the term $c^w_1$ of eq.(3.17), has been calculated in refs.[28,29]
with the help of the equivalence theorem. After transverse polarized
vector bosons are included in the final state, this term leads to
a very small effect for center of mass energy values up to 1 TeV.
Here we neglect this term and instead we calculate the leading
$\ln m^2$ contribution due to a heavy Higgs for the unpolarized
$\gamma\gamma\to W^+W^-$ cross section. According to the heavy
Higgs model the $\ln m^2$ term must be interpreted as an unknown
parameter.
\bigskip
\noindent{\bf 5.a\ \ The differential cross section in lowest non-zero order}
\bigskip
In lowest non-zero order in perturbation theory, the contributing
Feynman diagrams to the amplitude of the process
$\gamma\gamma\to W^+W^-$ are shown in
fig.9. For the corresponding expression of eq.(2.8), i.e.
$$A^{ww}(\lambda_1,\lambda_2,\lambda_3,\lambda_4)  =\epsa\epsb\epsc\epsd
\cdot A^{ww}_{\alpha\beta\mu\nu},\eqno(5.1)$$
we have
$$\eqalignno{ & A^{ww}_{\alpha\beta\mu\nu}  = 4\pi\alpha\cdot \left\{
-2\delta_{\mu\nu}\delta_{\alpha\beta}
-\delta_{\mu\alpha}\delta_{\nu\beta}\cdot {s\over p_1k_1}
-\delta_{\nu\alpha}\delta_{\mu\beta}\cdot {s\over p_1k_2}\right.\cr
& +(\delta_{\mu\alpha}k_{1\nu}p_{1\beta}
   +\delta_{\mu\beta}p_{1\alpha}k_{2\nu}
   -\delta_{\mu\nu}p_{1\alpha}p_{1\beta}
    - \delta_{\nu\alpha}k_{1\mu}p_{1\beta}
- \delta_{\nu\beta}p_{1\alpha}k_{2\mu})\cdot \left( {2\over p_1k_1}
+{2\over p_1k_2}\right)\cr
& +  \delta_{\alpha\beta}k_{1\mu}k_{2\nu}\cdot {2\over p_1k_1}
  + \left.
\delta_{\alpha\beta}k_{2\mu}k_{1\nu}\cdot{2\over p_1k_2}\right\}&(5.2)\cr}$$
The dotproducts are given by
$$p_1k_1=s/4\cdot(-1+\beta_w \cos\theta),\ \ \ \
p_1k_2=s/4\cdot(-1-\beta_w \cos\theta).\eqno(5.3)$$
When both outgoing vector bosons are longitudinally
polarized we have
$$\eqalignno{A^{ww}(+,+,0,0) & =
{4\pi\alpha\over 1-\beta_w^2\cos^2\theta}\cdot
{-8M^2\over s}\to {\cal O}\left( {M^2\over s}\right)
{\rm \ for\ large\ }s&(5.4)\cr
A^{ww}(+,-,0,0) & ={4\pi\alpha\over 1-\beta_w^2\cos^2\theta}
\cdot (1-\cos^2\theta)
\cdot \left( -2-{4M^2\over s}\right)\to {\cal O}(1)
{\rm \ for\ large\ }s.\ \ \ &(5.5)\cr}$$
In the limit $M\ll E$, $A^{ww}_{\alpha\beta\mu\nu}$ is of the order one
and when multiplied
with the longitudinal polarization vectors, we expect $A(+,+,0,0)$ and
$A(+,-,0,0)$ to contain ${\cal O}(E^2/M^2)$ terms. Due to the Yang-Mills
type cancellations these terms do not survive.
The unpolarized differential cross section is given by
$${d\sigma^{ww}\over d\cos\theta}={\beta_w\over 32\pi s}\cdot (4\pi\alpha)^2
\cdot\left\{ 6-{32+48M^2/s\over 1-\beta^2\cos\theta^2}
+{64+192M^4/s^2\over (1-\beta^2\cos\theta^2)^2}\right\},\eqno(5.6)$$
for which the longitudinal contribution of eq.(5.4) may be neglected
for energies already slightly above threshold. In fig.10
we show the unpolarized differential cross section as a function
of $\cos\theta$ for center of mass energy values $\sqrt{s}=200$ GeV,
400 GeV and 1 TeV.
\bigskip
\noindent{\bf 5.b\ \ The one loop correction due to a heavy Higgs}
\bigskip
Up to a center of mass energy of 1 TeV, the only heavy Higgs effects
that may possibly be observable are those that also exist in
the $W^+_TW^-_T$ channel. In this channel the leading heavy Higgs
effect is the $\ln m^2$ correction, while there is no bad high energy
behaviour. In accordance with the screening theorem the quadratic Higgs
mass dependence shows up at the two loop level.
Consider the amplitude for the process
$\gamma\gamma\to W^+W^-$ near threshold.
The leading heavy Higgs effect is given by eq.(3.19):
$$A^{ww}=\alpha\cdot c^w_a+\alpha\alpha_w\cdot
\left\{ c^w_b\cdot \ln {m^2\over M^2}+{\cal O}(1)\right\},\eqno(5.7)$$
We also could have considered the amplitude at some other energy up to 1 TeV.
The leading heavy Higgs effect would still be as given above,
while the energy dependence enters in the form of some  $\ln(p^2/M^2)$ term
($p$ is a typical momentum).
The one loop calculation that must be done to extract
the $\ln m^2$ term is a
standard one. As we mentioned in section 3.c we do not give the details
and we only give a sketch of how the $\ln m^2$ terms
are obtained. The receipe is as follows:
\item{(1)} leave the propagators and the external lines unchanged
\item{(2)} replace the occuring tree vertices in the diagrams of figs.
9a-9c, by the one loop corrected vertices
\item{(3)} evaluate the irreducible one loop $\gamma\gamma W^+W^-$ diagrams

\noindent Expressions for various one loop corrected vertices may be found
in appendix E of ref.[15].
Note that there are no $U$ particle effects, as they enter at the two loop
level. We start with the vector boson
exchange diagrams of fig.9a, which at the tree level may be written as
$$A_{\abcd}(a)=\left\{\Gamma^0_{\gamma ww}\cdot P_w\cdot
\Gamma^0_{\gamma ww}\right\}_{\abcd}. \eqno(5.8)$$
The notation should be clear: $\Gamma^0$ is the tree level vertex,
and $P$ is the propagator.
The heavy Higgs mass correction is found by replacing
$\Gamma^0_{\gamma ww}$ by $\Gamma^R_{\gamma ww}$, where $\Gamma^R$
is the one loop corrected
vertex, keeping only the $\ln m^2$ terms. It is found
that this correction is proportional to the tree level vertex:
$$\Gamma^R_{\gamma ww}=\Gamma^0_{\gamma ww}\cdot \delta_{\gamma ww},
\eqno(5.9)$$
with
$$\delta_{\gamma ww}=-\delta_g-{5\over 12}\cdot{g^2\over 16\pi^2}
\cdot \ln m^2.
\eqno(5.10)$$
Note that we left $\delta_g$ unspecified; possible values are given
in eqs.(3.20) and (3.21). The one loop correction to the diagram
of fig.9a is thus given by
$$C_{\abcd}(a)=A_{\abcd}(a)\cdot 2\delta_{\gamma ww},\eqno(5.11)$$
and is proportional to the tree level diagram.

Similarly, for the Higgs ghost exchange diagrams of fig.9b, given by
$$A_{\abcd}(b)=\left\{ \Gamma^0_{\gamma w \phi}\cdot P_{\phi}\cdot
\Gamma^0_{\gamma w\phi}\right\}_{\abcd}, \eqno(5.12)$$
we must replace the tree vertex $\Gamma^0_{\gamma w\phi}$ by
$\Gamma^R_{\gamma w\phi}$. Again the corrected vertex is proportional to the
tree level vertex, i.e.
$\Gamma^R_{\gamma w\phi}=\Gamma^0_{\gamma w\phi}\cdot \delta_{\gamma w\phi}$,
with
$$\eqalignno{
\delta_{\gamma w\phi} & =-\delta_g-{5\over 12}\cdot{g^2\over 16\pi^2}
\ln m^2
\cr
& =\delta_{\gamma ww}.&(5.13)\cr}$$
This result may be easily understood by considering the on-shell Ward
identity for the $\gamma WW$ vertex.
Thus
$$C_{\abcd}(b)=A_{\abcd}(b)\cdot 2\delta_{\gamma ww}.
\eqno(5.14)$$
As a result of the shifts of the fields and the parameters,
the $\gamma\gamma WW$ four-point vertex of fig.9c receives the
counter term:
$$\Gamma^c_{\gamma\gamma ww}=\Gamma^0_{\gamma\gamma ww}
\cdot \delta_{\gamma\gamma ww}, \eqno(5.15)$$
with
$$\eqalignno{\delta_{\gamma\gamma ww} & =-2\delta_g
-{3\over 4}\cdot{g^2\over 16\pi^2}\cdot \ln m^2
\cr
& =2\delta_{\gamma ww}+{1\over 12}\cdot{g^2\over 16\pi^2}
\cdot \ln m^2.&(5.16)\cr}$$
All that is left to be done is the evaluation of the irreducible $\gamma
\gamma WW$ diagrams. The diagrams that contribute to the leading $\ln m^2$
terms are shown in fig.11. The result is
$$\Gamma^1_{\gamma\gamma ww}=-\Gamma^0_{\gamma\gamma ww}
\cdot{1\over 12}\cdot {g^2\over 16\pi^2}\cdot \ln m^2.\eqno(5.17)$$
The one loop corrected $\gamma\gamma WW$ four point vertex
is thus given by
$$C_{\abcd}(c)=\{\Gamma^c_{\gamma \gamma ww}+
\Gamma^1_{\gamma\gamma ww}\}_{\abcd}=
\{\Gamma^0_{\gamma\gamma ww}\cdot
2\delta_{\gamma ww}\}_{\abcd}.\eqno(5.18)$$
When we add up all the contributions given by  eqs.(5.11), (5.14) and (5.18),
we find for the one loop corrected amplitude $A^{ww}_1=A^{ww}+C^{ww}$,
$$A^{ww}_1\lalblcld=A^{ww}\lalblcld\cdot(1+2\delta_{\gamma ww}).
\eqno(5.19)$$
This result may also be derived by considering on-shell Ward
identities, without actually performing the exact one loop
calculation for the $\gamma\gamma WW$ vertex.
{}From eq.(3.21) we observe that
the process $\gamma\gamma\to WW$ receives the same $\ln m^2$ correction
as Coulomb scattering.

The $\ln m^2$ term needs to be interpreted
as an unknown parameter and must be fixed by experiment.
According to the heavy Higgs model the parameter measured in
the process $\gamma\gamma\to WW$ at a center of mass energy less than 1 TeV,
is the same as the parameter measured in low energy processes like
the $\rho$ parameter, muon decay or Coulomb scattering.
A discrepancy would indicate that higher order corrections are important.
\bigskip
\bigskip
\noindent{\bf 6.\ \ The process $\gamma\gamma\to t\bar t$}
\bigskip
The top quark has yet to be discovered. From the CDF experiment at Fermilab,
the lower bound on the top quark mass is given by 100 GeV.
{}From the precision measurements performed at the LEP experiments, the
upperbound is derived to be 200 GeV when assuming the validity of
perturbation theory.

The Higgs exchange diagram of fig.1 with a $t\bar t$ pair in the final
state is a next-to-leading order contribution to
the amplitude for the process $\gamma\gamma
\to t\bar t$. Nevertheless it may lead to a significant effect since the
Higgs $t\bar t$ coupling is given by the Yukawa coupling and is proportional
to $m_t/M={\cal O}(1)$.
For a given finite Higgs mass value, this contribution has been
examined in ref.[30] for  the purpose of testing  the Yukawa coupling.
\bigskip
\noindent{\bf 6.a\ \ The amplitude and the cross-section in lowest non-zero
order}
\bigskip
The tree level Feynman diagram for the process $\gamma\gamma\to t\bar t$
is shown in fig.12. The corresponding expression for the amplitude in lowest
non-zero order in perturbation theory is given by
$$\eqalignno{& A^{t\bar t}(\la,\lb)=2\pi\alpha Q^2\cdot \epsa\epsb\cr
\cdot & \left\{
{1\over p_1k_1}\{ \bar u (p_2)\gamma^{\beta}
  [-i(\k_1-\p_1)+m_t]\gamma^{\alpha} u (-p_1)\}\right. \cr
& + \left. {1\over p_1k_2}\{\bar u (p_2)\gamma^{\alpha}
  [-i(\k_2-\p_1)+m_t]\gamma^{\beta} u (-p_1)\}\right\},&(6.1)\cr}$$
with $Q=2/3$, $p_1k_1=s/4\cdot(-1+\beta_t\cos\theta)$ and
$p_1k_2=s/4\cdot(-1-\beta_t\cos\theta)$. Using eq.(2.15), we find for
the $J=0$ differential cross-section
$${d\sigma^{t\bar t}_0(++)\over d\cos\theta}=
{16\pi\alpha^2 Q^4 N_c\beta_t \over (1-\beta_t^2\cos^2\theta)^2}\cdot
\left( {m_t^2\over s^2} \right) \cdot (1+\beta_t^2).\eqno(6.2)$$
When the initial photons are in the $J=2$ state, we have
$${d\sigma^{t\bar t}_0(+-)\over d\cos\theta}=
{4\pi\alpha^2 Q^4 N_c\beta_t \over s(1-\beta_t^2\cos^2\theta)^2}\cdot
\beta_t^2 \sin^2\theta\cdot (2-\beta^2_t\sin^2\theta),\eqno(6.3)$$
where $N_c=3$ and $1-\beta_t^2=4m_t^2/s$.
The subscript 0 indicates the lowest non-zero order contribution only.
At large scattering angles  the cross-section behaves like $1/s$
in the $J=2$ state and behaves like $m_t^2/s^2$ in the $J=0$ state.
The integrated cross-sections are found to be
$$\eqalignno{ & \sigma^{t\bar t}_0(++)=8\pi\alpha^2Q^4N_c\beta_t\cdot
{m_t^2\over s^2}\cdot(1+\beta_t^2)\cr
\cdot & \left\{ {2c_m\over 1-\beta_t^2c_m^2}+{1\over \beta_t}
\ln\left( {1+\beta_t c_m\over 1-\beta_t c_m}\right)\right\}
,&(6.4)\cr}$$ and
$$\eqalignno{ \sigma^{t\bar t}_0(+-)= & {8\pi\alpha^2 Q^4N_c\beta_t\over s}
\cdot\left\{ -c_m-{4m_t^2\over s}\left( 1+{2m_t^2\over s}\right)
\cdot {c_m\over 1-\beta_t^2c_m^2}\right.\cr
+ & \left. {1\over \beta_t}\left(1+{2m_t^2\over s}-{4m_t^4\over s^2}
\right)\ln \left( {1+\beta_t c_m\over 1-\beta_t c_m} \right) \right\},
&(6.5)\cr}$$
where $\cos\theta$ is integrated from  $-c_m$ to $c_m$.
\bigskip
\noindent{\bf 6.b Finite Higgs mass effects}
\bigskip
We only consider the next-to-leading order effects due to the Higgs exchange
diagram of fig.1. Although we neglect many other diagrams
we expect that near the resonance, i.e. when $-s+m^2={\cal O}(g^2)$,
it will give the leading correction for not too high Higgs mass values.
Furthermore, at $-s+m^2=0$, this diagram is gauge invariant.
The correction will only contribute when the initial photons have the same
helicity. As may be derived easily from the discussion of section 3.b,
the corresponding expression for the amplitude is given by
$$A^{t\bar t}_H(++)=\alpha \alpha_w\cdot { m_t\over 2 M^2}
\{\bar u(p_2)u(-p_1)\}\cdot {\cal H}(s),\eqno(6.6)$$
with
$${\cal H}(s)=P_H\cdot A(H).\eqno(6.7)$$
$P_H$ is given by eq.(3.12), and is gauge invariant because the width
is gauge invariant. $A(H)$, given by eq.(3.9),
is gauge invariant when evaluated at $-s+m^2=0$.
Including this correction, the differential cross-section may be written as
$${d\sigma(++)\over d\cos\theta}={d\sigma^{t\bar t}_0(++)\over d\cos\theta}+
{d\sigma^{t\bar t}_1(++)\over d\cos\theta}+
{d\sigma^{t\bar t}_2(++)\over d\cos\theta}.\eqno(6.8)$$
$d\sigma^{t\bar t}_0(++)/d\cos\theta$ is the cross-section
in lowest non-zero
order and is given by eq.(6.2). The expression for the interference term
$d\sigma^{t\bar t}_1(++)/d\cos\theta$ is given by
$${d\sigma^{t\bar t }_1(++)\over d\cos\theta}=
{\alpha^2\alpha_wQ^2N_c\beta_t\over 1-\beta_t^2 \cos^2\theta}\cdot
\left( {-\beta_t^2 m_t^2\over 2sM^2}\right)\cdot
\{{\cal H}(s)+{\cal H}^*(s)\}.\eqno(6.9)$$
The third and last term is of ${\cal O}(g^8)$ and clearly will only give
a significant contribution for a sufficiently
narrow width of the Higgs. We have
$${d\sigma^{t\bar t}_2(++)\over d\cos\theta}=
{\alpha^2\alpha^2_w N_c\beta_t\over 64\pi}\cdot
\left( {c_m\beta^2_t m_t^2\over M^4}\right)\cdot
\{{\cal H}(s)\cdot {\cal H}^*(s)\}.\eqno(6.10)$$
Writing ${\cal H}(s)=(a+ib)/\{-s+m^2-im\Gamma(s,g^2)\}$,
we have near the Higgs resonance
$${\cal H}(s)+{\cal H}^*(s)={-2b\over m\Gamma(g^2)}+{\cal O}(1),
\eqno(6.11)$$
and
$${\cal H}(s)\cdot {\cal H}^*(s)={(a^2+b^2)\over m^2\Gamma^2(g^2)}+
        {\cal O}\left( {1\over g^2} \right).\eqno(6.12)$$
Here
$$\eqalignno{ a & = 6M^2+m^2+6M^2\cdot (m^2-2M^2)\cdot{\cal R}e
\left( {C^w_0\over i\pi^2} \right)\cr
&  + Q^2N_cm_t^2\cdot \left\{ -4+(8m_t^2-2m^2)\cdot {\cal R}e
\left( {C^t_0\over i\pi^2} \right) \right\},&(6.13)\cr}$$
and
$$\eqalignno{b & = 6M^2\cdot (m^2-2M^2)\cdot {\cal I}m
\left( {C^w_0\over i\pi^2} \right)\cr
& +Q^2N_cm_t^2\cdot (8m_t^2-2m^2)\cdot
{\cal I}m\left( {C^t_0\over i\pi^2} \right).&(6.14)\cr}$$
The integrated cross-section is given by
$$\sigma^{t\bar t}(++)=\sigma^{t\bar t}_0(++)
+\sigma^{t\bar t}_1(++)+\sigma^{t\bar t}_2(++),\eqno(6.15)$$
with $\sigma^{t\bar t}_0(++)$ as given by eq.(6.4). Furthermore
$$\sigma^{t\bar t}_1(++)={\alpha^2\alpha_wQ^2N_c\beta_t\over s}\cdot
{m_t^2\over M^2}\cdot
\left\{ -{\beta_t\over 2}\ln\left( {1+\beta_t c_m\over 1-\beta_t c_m}
\right)\right\}\cdot\{{\cal H}(s)+{\cal H}^*(s)\}
,\eqno(6.16)$$
and
$$\sigma^{t\bar t}_2(++)={\alpha^2\alpha^2_wN_c\beta_t\over 32\pi}\cdot
{m_t^2\over M^4}\cdot\beta_t^2 c_m\cdot
 \{{\cal H}(s)\cdot {\cal H}^*(s)\}.\eqno(6.17)$$
\bigskip
\noindent{\bf 6.c The large Higgs mass limit}
\bigskip
We are interested in the contribution of the Higgs exchange diagram of fig.1
to the $\gamma\gamma\to t\bar t$ cross section in the limit
of a heavy Higgs.
It is derived from eq.(6.16) by taking
the limit of a large Higgs mass in the expression for ${\cal H}(s)$. Let us
write
$$\eqalignno{ {\cal H}_{\infty}(s) & =\{ {\cal H}(s)\}_{m\to \infty}\cr
& = 1 +{\cal O}\left( {M^2\over s} \right).&(6.18)\cr}$$
The $U$-particle contribution to ${\cal H}_{\infty}(s)$ is given by
$${\cal H}_U(s)=-{\alpha_w\over 4\pi c_w^2}\cdot \left(
{\beta^Us\over 8M^2}\right)\cdot\left\{ 1 + {\cal O} \left(
{M^2\over s} \right)\right\}.\eqno(6.19)$$
Including these corrections, we have
$$\sigma^{t\bar t}(++)=\sigma^{t\bar t}_0(++)+\sigma^{t\bar t}_{\infty}(++)
+\sigma^{t\bar t}_U(++),\eqno(6.20)$$
where $\sigma^{t\bar t}_0(++)$, given by eq.(6.4),
is the cross section in lowest non-zero order. The
corrections according to the heavy Higgs model are given by
$$\sigma^{t\bar t}_{\infty}(++)={\alpha^2\alpha_wQ^2N_c\beta_t\over s}
\cdot {m^2_t\over M^2}\cdot
\left\{ -{\beta_t\over 2}\ln\left( {1+\beta_t c_m\over 1-\beta_t c_m}
\right)\right\}\cdot
\{{\cal H}_{\infty}(s)+{\cal H}^*_{\infty}(s)\}\eqno(6.21)$$
and
$$\sigma^{t\bar t}_U(++)={\alpha^2\alpha_wQ^2N_c\beta_t\over s}
\cdot {m^2_t\over M^2}\cdot
\left\{ -{\beta_t\over 2}\ln\left( {1+\beta_t c_m\over 1-\beta_t c_m}
\right)\right\}\cdot
\{{\cal H}_U(s)+{\cal H}^*_U(s)\}.\eqno(6.22)$$
\bigskip
It is seen that the cross section for the $J=0$ state corresponds to
the following expansion in $s/M^2$:
$$\sigma^{t\bar t}(++)=\alpha^2\cdot {m_t^2\over s^2}\cdot \left\{ c^t_1 +
\alpha_w\cdot {s\over M^2}\cdot c_2^t+
\alpha^2_w\cdot {s^2\over M^2}\cdot
\left\{ c^t_3\cdot\ln {m^2\over s}+c^t_4+\ldots\right\}\right\},\eqno(6.23)$$
where we must make the replacement
$$c^t_4\to c^t_4+c^{tU}_4.\eqno(6.24)$$
{}From eqs.(6.4), (6.21) and (6.22) the expressions for
$c_1^t$, $c_2^t$ and $c_4^{tU}$ are found in a very straightforward way.
We therefore do not explictly write them down.
As it would require a two loop calculation,
we have not evaluated the expression for $c^t_3$.
Let us compare the above expression to the $s/M^2$ expansions for the
$\gamma\gamma\to Z_LZ_L$ and $\gamma\gamma\to W_LW_L$ cross sections. These
are obtained by squaring the amplitudes, given by eq.(3.17),
and subsequently multiplying with $1/s$.
We see that the $\gamma\gamma\to t\bar t$
process is suppressed by an overal factor $m_t^2/s$.
Thus although the heavy Higgs effects
are expected to be smaller in the $t\bar t$ channel than in the
$Z_LZ_L$ or $W_LW_L$ channel,
here the $t\bar t$ channel is relatively background free.

According to the heavy Higgs model the leading Higgs mass correction to
the $J=2$ cross section may be written as
$$\sigma^{t\bar t}(+-)=\alpha\cdot {1\over s}\cdot
\left\{ c^t_5+\alpha_w\cdot c^t_6\cdot \ln
{m^2\over M^2}+{\cal O}(1)\right\}.\eqno(6.25)$$
The term $c_5^t$ may be obtained by considering eq.(6.5).
The term $c^t_6$ is found by performing a one loop calculation in the
limit of a large Higgs mass.
This has been done in ref.[12] and the result is
$$\sigma^{t\bar t}(+-)=\sigma^{t\bar t}_0(+-)
\cdot\{ 1+4\delta_{\gamma ww}\}.\eqno(6.26)$$
The correction is identical to the correction for the unpolarized
$\gamma\gamma\to WW$ cross section.
\bigskip
\noindent{\bf 6.d \ \ Results}
\bigskip
In fig. 13 we show the total cross-section in lowest non-zero order,
with an angular cut of $c_m=0.7$ and for a top quark mass value of 150 GeV.
The $J=0$ cross-section is more rapidly vanishing for large
$s$ then the $J=2$ cross section.
\bigskip
In fig.14a we have plotted the effect due to the
Higgs exchange diagram of fig.1 for the two different models that are
considered in this paper. The top quark mass is taken to be 120 GeV.
The existence of the elementary Higgs boson as predicted by the
Standard Model Lagrangian
can lead to a significant effect,
notably for Higgs mass values below 500 GeV.
For Higgs mass values above 700 GeV the cross section becomes very
insensitive to the exact value of the Higgs mass.
On the other hand, there is a significant difference with respect to
the effect as predicted by the heavy Higgs model for $\beta^U \le 5$.
Assuming the validity of the partial wave analysis,
$\beta^U=5$ corresponds to a 2 TeV resonance in the $I=1$
channel for $W_LW_L$ scattering;
a smaller value for $\beta^U$ would indicate that the resonance is located
at a higher energy.

Figs.14b and 14c show similar curves for top quark mass values of
150 GeV and 180 GeV.
\bigskip
\bigskip
\noindent{\bf 7.\ \ Summary and concluding remarks}
\bigskip
For each of the processes $\gamma\gamma\to ZZ$, $\gamma\gamma\to WW$ and
$\gamma\gamma\to t\bar t$ we have investigated the level of sensitivity
to the Higgs sector of the Standard Model Lagrangian in the
energy region between 200 GeV and 1 TeV.
\bigskip
We have considered
the Standard Model Lagrangian containing the elementary Higgs boson
with a mass less than 1 TeV.
The elementary Higgs boson may be produced through $\gamma\gamma$ fusion,
given that the initial photons are in the $J=0$ state.
For Higgs mass values up to 400 GeV a resonance signal may be observed
in the $ZZ$ channel. This conclusion has also been reached in refs.[21-23].
For higher Higgs mass values the resonance signal
disappears and detecting the Higgs has turned into a precision measurement.
This is because the Higgs width has become very broad and furthermore
the transverse polarized vector boson production is a factor 100 larger
than the longitudinal vector boson production, and is a severe background.
For example for a Higgs mass value of 500 GeV there is a 10\%
enhancement. For even higher Higgs mass
values effects are less than $5\%$.
In the $WW$ channel the background due to transverse vector boson
production is even more substantial. In addition, the Higgs exchange
diagram of fig.1 contributes to the amplitude only in next-to-leading order.
For Higgs mass values higher than 300 GeV the effect is
less then a few percent [17].
In the $t\bar t$ channel the Higgs exchange diagram of fig.1
also contributes to the next-to-leading order only. However the $t\bar t$
channel is relatively background free.
For Higgs mass values above 400 GeV the size of the
effect in the $t\bar t$ channel
is about the same as the size of the effect in the $ZZ$ channel.
Also here, for Higgs masses larger than 600 GeV the effect is
reduced to less than a few percent.
\bigskip
If the elementary Higgs boson has not been found below an energy
of 1 TeV, we assume there is new physics which must show up in the
TeV region. To examine this possibility we have considered the
heavy Higgs model. This model corresponds to the
Standard Model Lagrangian plus an additional Higgs interaction
with a ficticious $U$ particle. The corresponding coupling strength
is denoted by $g_U$. The mass of the $U$ particle is taken to be
equal to the mass of the Higgs particle. Subsequently, after a typical
one loop calculation the limit of a heavy Higgs mass is taken.
In this way we are left with terms proportional to $\ln m^2$ and
$g_U^2$. These terms are to be interpreted as arbitrary parameters
as may be easily understood within the framework of effective
field theory. Assuming the validity of the one loop approximation,
the heavy Higgs model contains just these two arbitrary parameters.
If higher order corrections are important, the $\ln m^2$
and $g^2_U$ terms obtained for one process cannot be compared
to the $\ln m^2$ and $g_U^2$ terms obtained for another process:
they are then independent sets of unknown parameters.
Sensitivity to these arbitrary parameters implies sensitivity
to the new physics. A possible manifestation is the occurence of
a resonance in the $I=1$ channel for $W_LW_L$ scattering.
Assuming the validity of partial wave analysis, the location of this
resonance depends on the Lehmann $\beta$ parameter. The $\beta$ parameter
depends on $g_U^2$ only. If $\beta=5$ ($g^2_U=27$) the resonance is
located at 2 TeV, as predicted by QCD-like models.

We have evaluated the cross sections for the processes $\gamma\gamma\to ZZ$,
$\gamma\gamma\to WW$ and $\gamma\gamma\to t\bar t$, according to
the heavy Higgs model. We have examined the sensitivity that these
processes may have to the $\beta$ parameter.
Unfortunately, in the $ZZ$ and $WW$ channels effects are insignificant
due to the  the enormous $Z_TZ_T$ and $W_TW_T$ background.

The effects in the $t\bar t$ channel are more significant.
It may be possible to establish the non-existence of the elementary
Higgs boson with a mass of 1 TeV or less,
thus favouring the heavy Higgs model.
Sensitivity to the $\beta$
parameter, i.e. sensitivity to the new physics,
requires an ${\cal O}(1)$\% accuracy at $\sqrt{s}=1$ TeV.
At this energy the $\gamma\gamma\to t\bar t$ cross section
with the initial photons in the $J=0$ state
is about 50$-$100 fb (depending on the top quark mass).

We have not discussed the role of the $\ln m^2$ term in this paper. We only
wish to remark that this term, within the framework
of partial wave analysis,
may be correlated to a possible occurence of a resonance in the $I=0$
channel for $W_LW_L$ scattering [9-11].
\bigskip
\bigskip
\noindent{\bf References}
\bigskip
\+[1]&I. Ginzburg, G. Kotkin, V. Serbo and V. Telnov,\cr
\+&Nucl. Instrum. Methods 205 (1983) 47; 219 (1984) 5.\cr
\+[2]&C. Akerlof, Michigan preprint (1981) UMHE 81-59;\cr
\+&V. Telnov, Proc. of the conference on ``Physics and experiments\cr
\+&with linear colliders'' (1993), eds. R. Ovara, P. Eerola
and M. Nordberg\cr
\+&(Saariselka, Finland Sept. 9-14, 1991), and references therein.\cr
\+[3]&P. Pesic, Phys. Rev. D8 (1973) 945;\cr
\+&E. Yehudai, Phys. Rev. D44 (1991) 3434;\cr
\+&G. B\'elanger and F. Boudjema, Phys. Lett. B288 (1992) 210.\cr
\+[4]&D. Dicus and V. Mathur, Phys. Rev. D7 (1977) 3111\cr
\+[5]&B.W. Lee, C. Quigg and H. Thacker, Phys. Rev. D16 (1977) 1519;\cr
\+&G. Passarino, Nucl. Phys. B343 (1990) 31.\cr
\+[6]&M. Veltman, Acta Phys. Polon. B8 (1977) 475.\cr
\+[7]&R. Rosenfeld, Mod. Phys. Lett. A4 (1989) 19999;\cr
\+&M. Chanowitz and M. Golden, Phys. Rev. 61 (1988) 1053;\cr
\+&P. Hung and H. Thacker, Phys. Rev. D31 (1985) 2866;\cr
\+&P. Hung, T. Pham and T. Truong, Phys. Rev. Lett. 59 (1987) 2251;\cr
\+&R. Rosenfeld and J. Rosner, Phys. Rev. D38 (1988) 1530.\cr
\+[8]&M. Chanowitz and M.K. Gaillard, Nucl. Phys. B261 (1985) 379.\cr
\+[9]&H. Veltman and M. Veltman, Acta Phys. Polon. B (1992);\cr
\+&R.S. Willey, Phys. Rev. D44 (1991) 3646.\cr
\+[10]&A. Dobado, M.J. Herrero and T. Truong, Phys. Lett. B235
(1990) 129.\cr
\+[11]&H. Lehmann, Acta Phys. Austracia, Suppl. XI (1973) 139.\cr
\+[12]&J.J. van der Bij and M. Veltman, Nucl. Phys. B231 (1984) 205.\cr
\+[13]&M. Veltman and F. Yndurain, Nucl. Phys. B325 (1989) 1.\cr
\+[14]&M. Lemoine and M. Veltman, Nucl. Phys. B 164 (1980) 446;\cr
\+&R. Philippe, Phys. Rev. D26 (1982) 1588;\cr
\+&M. B\"ohm, A. Denner, T. Sack, W. Beenakker, F. Berends, and H.Kuijf,\cr
\+&Nucl. Phys. B304 (1988) 463;\cr
\+&J. Fleischer, F. Jegerlehner and M. Zralek, Z. Phys.
C42 (1989) 409.\cr
\+[15]&H. Veltman, Phys. Rev. D43 (1991) 2236.\cr
\+[16]&V. Jain, LAPP preprint, LAPP-TH-297/90.\cr
\+[17]&D. Morris, T. Truong and D. Zappal\'a, UCLA preprint,
UCLA/TEP/93/35.\cr
\+[18]&G. Valencia and S. Willenbrock, Phys. Rev. D46 (1992) 2247.\cr
\+[19]&T. Appelquist and C. Bernard, Phys. Rev. D22 (1980) 200.\cr
\+[20]&S. Kyriazidou, Michigan preprint, UM-TH-92-22.\cr
\+[21]&G. Jikia, Phys. Lett. B298 (1993) 224.\cr
\+[22]&M. Berger, Madison preprint, July 1993, MAD-PH-771.\cr
\+[23]&D. Dicus and C. Kao, Florida preprint, August 1993, FSU-HEP-93808.\cr
\+[24]&E. Glover and J.J. van der Bij, Nucl. Phys. B321 (1989) 561.\cr
\+[25]&M. Veltman, Schoonschip, a symbolic manipulator program.\cr
\+[26]&M. Veltman, Formf, a program that calculates one loop form factors.\cr
\+[27]&J. Cornwall, D. Levin, and G. Tiktopoulos, Phys. Rev. D10 (1974) 1145;
\cr
\+&B.W. Lee, C. Quigg, and H. Thacker, Phys. Rev. D16 (1977) 1519;\cr
\+&G. Gounaris, R. K\"ogerler, and H. Neufeld, Phys. Rev. D34 (1986) 3257;\cr
\+&H. Veltman, Phys. Rev. D41 (1990) 2294;\cr
\+&G. Valencia and S. Willenbrock, Phys. Rev. D42 (1990) 853.\cr
\+[28]&M. Herrero and E. Ruiz-Morales, Phys. Lett. B296 (1992) 397.\cr
\+[29]&E. Boos and G. Jikia, Phys. Lett. B275 (1992) 164.\cr
\+[30]&E. Boos, I. Ginzburg, K. Melnikov, T. Sack and
S. Shichanin,\cr
\+& Z. Phys. C56 (1992) 487.\cr
\bigskip
\bigskip
 \noindent{\bf Figure Captions}
 \bigskip
 \+Fig.1&Higgs production through $\gamma\gamma$ fusion.\cr
 \+Fig.2&Kinematics for the process $\gamma\gamma\to X\bar X$;
$X\bar X$ represents a $ZZ$, a $W^+W^-$ or a\cr
\+& $t\bar t$ pair.\cr
 \+Fig.3&$\gamma\gamma\to ZZ$ lowest order Feynman diagrams\cr
 \+&(when labelling the external lines, include crossings).\cr
 \+&The dashed lines represent charged W and charged
Higgs ghost $\phi$ lines.\cr
\+& The dotted lines represent, in addition to $W$ and $\phi$,
Faddeev-Popov ghost $\psi$\cr
\+&and fermion $f$ lines. The $\psi$ and $f$ lines have a
direction and\cr
\+& need to be supplemented with an arrow.\cr
\+Fig.4&The process $\gamma\gamma\to ZZ$ at a
 center of mass energy $\sqrt{s}=200$ GeV.\cr
\+4a&$J=0$ differential cross section as a function of $\cos\theta$.\cr
\+4b&$J=0:$ Ratio $R(m,\theta,\sqrt{s}=200\ {\rm GeV})$ as a function
of $\cos\theta$ in \%. \cr
\+&The value of the Higgs mass is as indicated.\cr
\+&$R(m,\theta,\sqrt{s})$ is defined in eq.(4.18).\cr
 \+4c&$J=2$ differential cross section as a function of $\cos\theta$.\cr
 \+Fig.5&Same as fig.4, but here at $\sqrt{s}=300$ GeV.\cr
 \+&The value for the top quark mass is as indicated.\cr
 \+Fig.6&Same as fig.4, but here at $\sqrt{s}=500$ GeV.\cr
 \+&The value for the top quark mass is as indicated.\cr
 \+Fig.7&Same as fig.4, but here at $\sqrt{s}=1$ TeV.\cr
 \+&The value for the top quark mass is as indicated.\cr
 \+Fig.8&$U$ particle correction to the $J=0$ differential cross section
 for the process\cr
\+& $\gamma\gamma\to Z_LZ_L$.
 The correction, in \%, is plotted as a function of $\cos\theta$\cr
\+& for $\beta^U=5$.\cr
 \+Fig.9&$\gamma\gamma\to W^+W^-$ lowest order Feynman diagrams\cr
 \+&(when labelling the external lines, include crossings).\cr
 \+Fig.10&The unpolarized differential cross section for the
 process $\gamma\gamma\to W^+W^-$\cr
\+& as a function as $\cos\theta$.\cr
 \+Fig.11&One loop Feynman diagrams that give a leading $\ln(m^2)$\cr
 \+&contribution to the process $\gamma\gamma\to W^+W^-$\cr
 \+&(when labelling the external lines, include crossings).\cr
 \+Fig.12&$\gamma\gamma\to t\bar t$ lowest order Feynman diagram\cr
 \+&(when labelling the external lines, include crossings).\cr
 \+Fig.13&The process $\gamma\gamma\to t\bar t$ for $m_t=150$ GeV.\cr
 \+&The $J=0$ and $J=2$ total cross sections are shown as a function
 of the\cr
\+& center of mass energy $\sqrt{s}$.
An angular cut of $c_m=0.7$ is applied.\cr
 \+Fig.14&Next-to-leading order correction in \% to the
$\gamma\gamma\to t\bar t$ cross section.\cr
\+14a&$m_t$ is taken to be 120 GeV.
The different curves show the effects according\cr
\+&to the different models considered in this paper.\cr
 \+14b&Same as fig.14a but here $m_t=150$ GeV.\cr
 \+14c&Same as fig.14a but here $m_t=180$ GeV.\cr
 \end

\end

\end

\end

\end

\end